\documentstyle[12pt,amssymb]{article}
\date{}
\begin{document}
\def\be{\begin{equation}}
\def\ee{\end{equation}}
\def\R{\Bbb R}
\def\C{\Bbb C}
\def\Z{\Bbb Z}
\def\d{\partial}

\def\la#1{\lambda_{#1}}
\def\teet#1#2{\theta [\eta _{#1}] (#2)}
\def\tede#1{\theta [\delta](#1)}
\def\N{{\frak N}}

\renewcommand{\theequation}{\thesection.\arabic{equation}}

\font\frak=eufm10 scaled\magstep1

\def\bra#1{\langle#1|}
\def\ket#1{|#1\rangle}
\def\goth #1{\hbox{{\frak #1}}}
\def\<#1>{\langle#1\rangle}
\def\cotg{\mathop{\rm cotg}\nolimits}

\centerline{\Large \bf A few remarks on integral
representation}

\medskip

\centerline{\Large \bf for zonal spherical functions}
 
 \medskip
 
\centerline{\Large \bf on the symmetric space $SU(N)/SO(N,\R)$.} 

\vskip 2cm

\centerline{ \large {\sc J.F. Cari\~nena
  and  A.M. Perelomov\footnote{On leave of absence from
Institute for Theoretical and Experimental Physics, 117259 Moscow, Russia} }}

\vskip 1cm

\centerline{Depto. F\'{\i}sica Te\'orica, Univ. de Zaragoza, 50009 
Zaragoza,
Spain.}

\vskip 2cm

\centerline{\sc Abstract}

{\small The integral representation on the orthogonal groups for
zonal spherical functions on the symmetric space $SU(N)/SO(N,\R)$
is used to obtain a generating function for such functions. For the case
$N=3$ the three--dimensional integral representation reduces to a
one--dimensional one.
 }

\medskip \medskip \medskip
\section{Introduction}

The interest of studying classical and quantum integrable
systems is always increasing. These systems present some very nice
characteristics which are  related to different algebraic and analytic 
properties. For instance, the connection of completely integrable
classical Hamiltonian systems  with semisimple Lie algebras
was established more than twenty years ago in \cite{OPIM}
and the relationship with quantum systems in \cite{OPLMP}.

On the other side, it was also shown in \cite{OPFA} and
\cite{OPPR} the possibility of finding the explicit form of the
Laplace--Beltrami operator for each symmetric space appearing in
the classification given in the  classical Helgason's book \cite{Hel}
by associating to it a quantum mechanical problem.

The search for the eigenfunctions of such operators is not an easy task.
These functions are but the so--called zonal spherical functions and
for one special case and for the case of symmetric spaces with root systems of
the type $A_{N-1}$ were found explicitly   in \cite{Pr}.

Our aim in this letter is to present some remarks concerning
the integral representation for zonal spherical functions
on the symmetric space $SU(N)/SO(N,\R)$. This  representation
will be used for
obtaining a generating function for such zonal spherical functions.

We recall that if $G$ is a connected real semisimple Lie group and
$T^\rho$ denotes an irreducible unitary representation of $G$
with support in the Hilbert space $\cal H$, where $\rho$ is a parameter
characterizing the representation, the representation $T^\rho$ is
said to be
of class I if there exists a vector $\ket{\Psi_0}$ such that
$T^\rho(k)\ket{\Psi_0}=\ket{\Psi_0}$,  for any element $k$ in the maximal
compact subgroup $K$ of $G$. The function defined by the
expectation value of $T^\rho$ is called a zonal spherical function
belonging to the representation $T^\rho$. Zonal spherical functions satisfy a
kind of completeness condition like  that of coherent states.

The paper is organized as follows. In order to the paper to be more
selfcontained  we give in Section 2  the general definitions and
properties on
zonal spherical functions. The particular case $N=2$ is considered
in Section 3,
and then the formulae are extended in Section 4 to the case $N=3$.
Section 5 is devoted to introduce
an integral representation for the generating
 function for zonal spherical functions for the symmetric space
 $SU(N)/SO(N,\R)$ and the integrals arising in the expresion are explicitly
computed in the particular cases $N=2$ and $N=3$.

\section{Zonal spherical functions}
\setcounter{equation}{0}

Let $G^-=SL(N,\R)$ be the group of real matrices of order $N$ with determinant
equal to one. This group
contains three important subgroups, to be denoted  $K,\, A$ and $\cal N$.
The subgroup $K=SO(N,\R)$ is the compact group of real
orthogonal matrices, the subgroup $A$ is the Abelian group of inversible
real diagonal
matrices and 
$\cal N$ is the subgroup of  lower triangular real matrices
with units on the principal diagonal,  which is a nilpotent group.

Using the polar decomposition of a matrix, the homogeneous  
space $X^{-}=G^{-}/K$ can be identified with  the space of real
positive--definite symmetric matrices with determinant equal to one. It is 
known that any element $g\in G^-$ may be decomposed in a unique way
as a product $g=kan$,
with $k\in K$, $a\in A$ and $n\in {\cal N}$, respectively, so-called 
Iwasawa decomposition. We denote the elements in such a factorization
 as $k(g),\, a(g)$
 and $n(g)$, i.e. $g=k(g)\, a(g)\, n(g)$. Correspondingly, the linear
space underlying the Lie algebra
${\goth g}$ of $G^-$ can be decomposed as a direct sum of
the  linear spaces of the Lie subalgebras ${\goth k}$ of $K$,
${\goth a}$ of $A$ and ${\goth n}$ of ${\cal N}$, i.e.,
${\goth g}={\goth k}\oplus{\goth a}\oplus{\goth n}$.  Let us also denote as
${\goth a}^*$ the dual space of ${\goth a}$ and  so on.

There are natural left and right actions of group $G^-$ on $K$ 
and ${\cal N}$, respectively,  induced by left and right multiplication, respectively,
which are defined by the formulae
\be k^g=k(gk),\quad n_g=n(ng),\ee
and for any $\lambda\in {\goth a}^*$, we may construct the
representation $T^\lambda (g)$ of the  group $G^-$ in the 
space of $L^2(K)$ or $L^2({\cal N})$ of square integrable functions on $K$ or ${\cal N}$ 
by the formula
\be [T^\lambda (g)\,f](k)=\exp \,\big ( (i\lambda -\rho ,\,H(gk)\big )\,
f(k^{g^{-1}}),\ee
or
\be [T^\lambda (g)\,f](n)=\exp \,\big ( (i\lambda -\rho ,\,H(ng)\big )\,
f(n_g),
\ee
where $H(g)$ is defined by $a(g)=\exp \,H(g)$ and $\rho$ is given by one
 half of the sum of positive roots of the symmetric space $X^-$, 
$$\rho=\frac 12\sum_{R^+}\alpha.
$$

This so called representation of principal series is unitary and irreducible. 
It has the property that in the Hilbert space ${\cal H}^\lambda $ there is 
a normalized vector $\ket{\Psi _0}\in {\cal H}^\lambda $
which is invariant under  the action of group $K$:
\be T^\lambda (k)\,\ket{\Psi _0}=|\Psi _0\rangle ,\ee
Let us consider the function
\be \Phi _\lambda (g)=\bra{ \Psi _0}\,T^\lambda (g)\,\ket{\Psi _0} .
\ee
This function is called a zonal spherical function and has the properties
 of 
\be \Phi _\lambda (k_1gk_2)=\Phi _\lambda (g),\   \Phi _\lambda (k)=1,\
 \forall k\in K,\ 
 \Phi _\lambda (e)=1.\label{propzsf}
\ee

For the realization of ${\cal H}^\lambda $ as $L^2(K)$, we take
$\ket{\Psi _0}$ as the constant function 
$\Psi _0(k)\equiv 1$, and then we have the integral representation for 
$\Phi _\lambda  (g)$:
\be \Phi _\lambda (g)=\int _K\,\exp \,\big ( (i\lambda -\rho ,\,H(gk)\big )\,
d\mu (k),\quad \int _K\,d\mu (k)=1,\label{intrep}
\ee
where $d\mu (k)$ denotes an invariant (under $G^-$) measure on $K$.
Note that due to (\ref{propzsf}) the function $\Phi _\lambda (g)$ is
completely  defined by the values
$\Phi _\lambda (a),\,a\in A$.

Here $\Phi _\lambda (g)$ is the eigenfunction of Laplace-Beltrami $\Delta _j$ 
operators and correspondingly $\Phi _\lambda (a)$ is the eigenfunction of 
radial parts $\Delta _j^0$ of these operators, in particular, 
\be \Delta _2^0=\sum _{j=1}^N\,\partial _j^2+2\kappa \sum _{j<k}^N\,\coth (q_j-q_k)
(\partial _j-\partial _k),\ \kappa=\frac 12 ,\ \partial _j=\frac{\partial }{\partial q_j},
\ a_j=e^{q_j}. \ee

Note that the analogous consideration of groups $G^-=SL(N,{\Bbb C})$
and $G^-=SL(N,{\Bbb H})$ over complex numbers and quaternions gives us the corresponding 
integral representations for $\kappa =1$ and $\kappa =2$.

Note that the above construction is also valid for the  dual spaces
$X^+=G^+/K$, where $G^+=SU(N)$ is the group of unitary matrices with
determinant equal to one. In this case the representation $T^\lambda (g)$
is defined by a set $l=(l_1,\ldots,l_{N-1})$ of $(N-1)$ nonnegative
integer numbers $l_j$ and the integral
 representation (\ref{intrep})
takes the form
\be
\Phi_l(g)=\int_K\exp(l,H(gk))\, d\mu (k),\quad \int_K d\mu (k) =1,
\label{intrepsu}
\ee
and $\Phi_l(g)$ is the eigenfunction of the radial part of the
Laplace--Beltrami operator
\be
\Delta _2^0=\sum _{j=1}^N\,\partial _j^2+2\kappa \sum _{j<k}^N\,
\cotg (q_j-q_k)
(\partial _j-\partial _k),\, \kappa=\frac 12 ,\,\partial _j
=\frac{\partial }{\partial q_j},
\, a_j=x_j=e^{i q_j}.
\ee

The element $k$ of the group $SO(N ,\R)$ is the matrix $(k_{ij})$ and
may be considered as
the set of $N$ unit orthogonal vectors $k^{(j)}=(k_{1j},\ldots,k_{Nj})$
from which we may construct the set of polyvectors
\be
k^{(i)}, \ k^{(i_1,i_2)}=k^{(i_1)}\wedge k^{(i_2)}, \
 k^{(i_1,i_2,i_3)}=k^{(i_1)}\wedge k^{(i_2)}\wedge k^{(i_3)},\ \cdots\ .
\ee

There is a natural action of the group $G$ on the space of polyvectors
and the integral representation (\ref{intrepsu}) may be written now in the form
\be
\Phi_{ l}(x_1,\ldots,x_N)=\int \Xi_1^{l_1}(x;k)\cdots
\Xi_{N-1}^{l_{N-1}}(x;k)\,d\mu(k^{(1)},\ldots,k^{(N-1)}),
\ee
where
\begin{eqnarray}
&&\Xi_1(x;k)=\sum_j k_j^{(1)}\,^2x_j, \
\Xi_2(x;k)=\sum_{i<j} (k^{(1)}\wedge k^{(2)})^2_{ij} \,x_ix_j,\nonumber\cr
&&\Xi_3(x;k)=\sum_{i<j<l} (k^{(1)}\wedge k^{(2)}\wedge k^{(3)})^2_{ijl}
 \,x_ix_jx_l,\ldots\ .
\end{eqnarray}

Here $d\mu(k^{(1)},\ldots,k^{(N-1)})$ is the invariant
measure on $K$ such that
\be
\int_K d\mu (k^{(1)},\ldots,k^{(N-1)}) =1.
\ee

\section{The case  $N=2$}
\setcounter{equation}{0}

In this case, the integral representation takes the form
\be \Phi_l(x_1,x_2)=\int \, [(k'ak)_{11}]^l\,d\mu (k)=\int \,(k_{11}^2x_1
+k_{21}^2x_2)^l\,
d\mu (k),\quad \int \,d\mu (k)=1,\ee
where $k'$ is the transpose matrix of $k$, or
\be \Phi_l(x_1,x_2)=\int _{S^1}\,(n_1^2x_1+n_2^2x_2)^l\,d\mu (n),\quad 
(n,n)=n_1^2+n_2^2=1,\ee
where $d\mu (n)=\frac{1}{2\pi }\,d\varphi $ is an invariant measure on an 
unit circle $S^1$ in ${\Bbb R}^2$.

So,
\be \Phi_l(x_1,x_2)=\sum _{k_1+k_2=l}C_{k_1,k_2}^lx_1^{k_1}x_2^{k_2},\ee
and
\be 
C_{k_1,k_2}^l = \frac{l!}{k_1!k_2!}\,\langle n_1^{2k_1}n_2^{2k_2}\rangle ,
 \quad
\langle n_1^{2k_1}n_2^{2k_2}\rangle = \int _{S^1}\,n_1^{2k_1}n_2^{2k_2}\,
d\mu (n).
\ee
The integral is easily  calculated by using a standard parametrization $n_1=
\cos \,\varphi ,\,\,n_2=\sin \,\varphi ,\,\,d\mu (n)=\frac{1}{2\pi }\,
d\varphi $. We obtain
\be \langle n_1^{2k_1}n_2^{2k_2}\rangle =\frac{(\frac{1}{2})_{k_1}\,(\frac
{1}{2})_{k_2}}{(1)_{k_1+k_2}},
\ee
where $(a)_k$ is the Pochhammer symbol $(a)_k=a(a+1)\cdots (a+k-1)$.
So finally we have
\begin{eqnarray} 
C_{k_1k_2}^l&=&\frac{(\frac{1}{2})_{k_1}\,(\frac{1}{2})_{k_2}}{(1)_{k_1}\,
(1)_{k_2}},\quad l=k_1+k_2,\\
\Phi_l(x_1,x_2)&=&\sum _{k_1+k_2=l}\frac{(\frac{1}{2})_{k_1}\,(\frac{1}{2})_
{k_2}}{(1)_{k_1}\,(1)_{k_2}\,}\,x_1^{k_1}x_2^{k_2}.\label{phil}
\end{eqnarray}
If we put $x_1=e^{i\theta },\,\,x_2=e^{-i\theta }$, then $\Phi_l(x_1,x_2)=A_lP_l
\,(\cos \,\theta )$, where $P_l(\cos \,x)$ is the Legendre polynomial.

These formulae may be easily extended to the $N$-dimensional case. Namely,
 we have 
\be \Phi_{(l,0,\ldots,0)}(x_1,\ldots ,x_N)=\int _{S^{N-1}}(n_1^2x_1+\ldots +n_N^2x_N)^l\,
d\mu (n),\quad \int \,d\mu (n)=1,\ee 
where $d\mu (n)$ is invariant measure on $S^{N-1}$ and
\begin{eqnarray} 
\Phi_{(l,0,\ldots,0)}(x_1,\ldots ,x_N)&=&\sum _{k_1+\ldots +k_N=l}C_{k_1\ldots k_N}^l\,
x_1^{k_1}\ldots x_N^{k_N},\nonumber \\
C_{k_1,\ldots k_N}^l &=& \frac{l!}{k_1!\ldots k_N!}\,
\langle n_1^{2k_1}\ldots 
n_N^{2k_N}\rangle ,\\
\langle n_1^{2k_1}\ldots n_N^{2k_N}\rangle &=& \frac{(\frac{1}{2})_{k_1}
\ldots (\frac{1}{2})_{k_N}}{(\frac{N}{2})_l}.\nonumber 
\end{eqnarray}
So
\be C_{k_1\ldots k_N}^l=\frac{(\frac{1}{2})_{k_1}\ldots (\frac{1}{2})_{k_N}}
{(1)_{k_1}\,\ldots (1)_{k_N}}\, \frac{(1)_{l}}{(\frac{N}{2})_{l}},\quad l=k_1+\ldots +k_N.\ee

\section{The case $N=3$}
\setcounter{equation}{0}

In this case, the element of the orthogonal group $SO(3,\R)$ has the form
\begin{displaymath}
\mbox{k}=\left( \begin{array}{ccc}
n_1&l_1&m_1\\
n_2&l_2&m_2\\
n_3&l_3&m_3
\end{array} \right), 
\end{displaymath}
i.e., it may be represented by the three unit orthogonal each other vectors 
$$  n,\, l,\, m;\,\, n^2= l^2= m^2=1,\,\, (n, l)=
(l,m)=( m, n)=0, $$
and the integral representation for zonal spherical polynomials takes
the form
\be \Phi_{pq}(x)=\int _K\,(n_1^2x_1+n_2^2x_2+n_3^2x_3)^p\left(\sum_{j<k}
(n_jl_k-
n_kl_j)^2\,x_jx_k\right)^q\,d\mu (n,l),\ee
where the integration is taken on the orthogonal group $K=SO(3,\R)$, what is 
equivalent to the space of two unit orthogonal vectors $n$ and $ l$.

Note that $m_k=\epsilon_{kij} n_il_j$; we also have  $x_1x_2=x_3^{-1}$,\ldots. 
Hence,
\be
\Phi_{pq}(x_1,x_2,x_3)=\int _K\,(n_1^2x_1+n_2^2x_2+n_3^2x_3)^p
\left(m_1^2x^{-1}_1+m_2^2x^{-1}_2+m_3^2x^{-1}_3\right)^q\, d\mu(n,m).
\ee
For vectors $n$ and $m$ the standard parametrization through
Euler angles $\varphi, \theta$ and $\psi$, may be used:
\begin{eqnarray}
n&=&(\cos \,\varphi \,\,\sin \,\theta ,\, \sin \,\varphi \,\,\sin \,
\theta,\,\cos \,\theta) ,\quad m=\cos\psi \cdot a+\sin\psi\cdot  b,\nonumber \\
a&=&(-\sin \,\varphi,\cos \,\varphi,0),\ b=(-\cos \,\varphi \,\,\cos \,\theta ,
-\sin \,\varphi \,\,\cos \,\theta,\sin\theta)
\end{eqnarray}
with $d\mu (k)=d\mu (n,m)=A\, \sin \,\theta \,d\theta \,d\varphi \,d\psi $,
and in the preceding expression we have a three--dimensional integral 
which may be calculated using the generating functions.

\section{Generating functions}
\setcounter{equation}{0}
Let us define the generating function by the formula
\be  F(x_1,x_2,\ldots ,x_N;t_1,\ldots,t_{N-1})=\sum \,
\Phi_{l_1\cdots l_{N-1}}(x_1,\ldots ,x_N)
\,t_1^{l_1}\cdots t_{N-1}^{l_{N-1}}.
\ee

Then we have the integral representation
\begin{equation}
F(x_1,x_2,\ldots ,x_N;t_1,\ldots,t_{N-1})=\int [\prod_{j=1}^{N-1}
(1-\Xi_j(x;k)t_j) ]^{-1}\,
d\mu(k). \label {integrep}
\end{equation}

Let us introduce the coordinate system such that $a$ and $b$ are two
unit orthogonal vectors in the two--dimensional plane orthogonal to the set of
vectors
$\{k^{(1)},\ldots,k^{(N-2)}\}$. Then, an arbitrary  unit vector $n$
 in this plane
has the form $\cos\psi \cdot a+ \sin \psi\cdot b$, and we may integrate
first on  $d\mu(n)$. The integral representation (\ref{integrep})
takes the form:
\begin{equation}
F(x_1,x_2,\ldots ,x_N;t_1,\ldots,t_{N-1})=\int [A_{ij}n_in_j]^{-1}\,
d\mu^{(N-2)}(k)\, d\mu(n).
\end{equation}

The integral on $d\mu(n)$ may be easily calculated  and we have 
\be
F(x_1,x_2,\ldots ,x_N;t_1,\ldots,t_{N-1})=\int [D]^{-1/2}\,
d\mu(k^{(1)},\ldots,k^{(N-2)}),
\ee
where $D=\det (A_{ij}), \, A_{ij}=A_{ij}(x;k^{(1)},\ldots,k^{(N-2)})$.

In the simplest case $N=2$, we have
\be
F(x_1,x_2;t)=[(1-x_1t)(1-x_2t)]^{-1/2},
\ee
from which the formula (\ref{phil}) for $\Phi_l(x_1,x_2)$ follows.

In the case $N=3$, the integration on $d\mu(\psi)$ gives
\begin{equation}
F(x_1,x_2,x_3;t_1,t_2)=\int B^{-1}(n)C^{-1/2}(n)\, d\mu(n), \
\int d\mu(n)=1,
\ee
where
\begin{equation}
B=1-(n_1^2x_1+n_2^2x_2+n_3^2x_3)t_1,\quad
C=(1-x_2^{-1}t_2)(1-x_3^{-1}t_2)n_1^2+\cdots
\end{equation}

The crucial step for further integration is  the use of the formula
\be
B^{-1} C^{-1/2}=\int_0^1 {d\xi}\,[B(1-\xi^2)+C\xi^2]^{-3/2}.
\ee

Using this formula we obtain 
\be
F(x_1,x_2,x_3;t_1,t_2)=\int_0^1d\xi\, \int
[E(x_1,x_2,x_3;t_1,t_2,n,\xi)]^{-3/2}\, d\mu(n),
\ee
where 
\be
E(x_1,x_2,x_3;t_1,t_2,n,\xi)=\sum_je_j(x_1,x_2,x_3;t_1,t_2,\xi) n_j^2.
\ee

We can now  integrate on $d\mu(n)$ and finally we obtain the
one--dimensional integral representation for the generating function
\be
F(x_1,x_2,x_3;t_1,t_2)=\int_0^1d\xi \,[H(x_1,x_2,x_3;t_1,t_2,\xi)]^{-1/2},
\label{gfir}
\ee
where $H=e_1e_2e_3$ and the functions $e_j(\xi;t_1,t_2)$ are given by
\be
h_j(\xi;t_1,t_2)=1-d_j(t_1,t_2)(1-\xi^2),\quad 
d_j(t_1,t_2)=(x_jt_1+x_j^{-1}t_2-t_1t_2).
\ee

From this it follows that if $z_1=x_1+x_2+x_3$, and 
$z_2=x_1x_2+x_2x_3+x_3x_1$, then
\begin{eqnarray}
H&=&a_0^3-a_0^2[z_1\tau_1+z_2\tau_2]+a_0[z_2\tau_1^2+z_1\tau_2^2+(z_1z_2-3)
\tau_1\tau_2]\cr
&-&[\tau_1^3+\tau_2^3+\tau_1\tau_2[(z_2^2-2z_1)\tau_1+(z_1^2-2z_2)\tau_2]]
\label{defH}
\end{eqnarray}
where $a_0=1+(1-\xi^2)t_1t_2$, $\tau_1=(1-\xi^2)t_1$, $\tau_2=(1-\xi^2)
t_2$. Note that from (\ref{defH}) it follows that the integral
(\ref{gfir}) is
elliptic and it may be expressed in terms of standard elliptic integrals.

Expanding $F(x_1,x_2,x_3;t_1,t_2)$ in power series of the variable $t_2$
one obtains
\be
F(x_1,x_2,x_3;t_1,t_2)=\sum_{q=0}^\infty F_q(x_1,x_2,x_3;t_1)\,t_2^q
\ee
and we have
\begin{equation}
 F_0(x_1,x_2,x_3;t)=\int_0^1d\xi\, [H_0]^{-1/2},
\end{equation} 
and
\be
F_1=\frac 12 \int_0^1d\xi\,H_1[H_0]^{-3/2}
\ee
where 
\begin{eqnarray*}
H_0&=&1-z_1\tau_1 +z_2\tau_1^2-\tau_1^3,\\
H_1&=&(1-\xi^2)z_2-[3\xi^2+z_1z_2(1-\xi^2)]\tau_1+[2z_1\xi^2+(1-\xi^2)z_2^2]
\tau_1^2-z_2\tau_1^3.
\end{eqnarray*}

From the integral representation (\ref{gfir}) many useful formulae
may be obtained, here we give just one of them: when $z_1$ and $z_2$ go  to
infinity,
\be
\Phi_{pq}(z_1,z_2)\approx A_{pq}\, z_1^pz_2^q, \quad
A_{pq}=\frac{(\frac 12)_{p}(\frac 12)_{q}}{(1)_{p}(1)_{q}}\,\frac{(1)_{p+q}}{(\frac 32)_{p+q}}
\ee

A more detailed version of this note will be published elsewhere.

\bigskip

{\parindent 0cm{\bf Acknowledgements.}} One of the authors (A.P.)
would like to thank the Department of Theoretical Physics of Zaragoza
 University for its hospitality. Financial support from Direcci\'on
General de Ense\~nanza Superior, Spain (SAB95-0447) is also acknowledged.

\end{document}